\documentclass{iau}
\usepackage{graphicx}
\usepackage{natbib}

\title[Asteroseismic Signatures of Magnetic Variation]
{Asteroseismic Signatures of Magnetic Activity Variations in Solar-type Stars}

\author[T.~S. Metcalfe]{Travis S. Metcalfe$^{1,2}$}

\affiliation{$^1$Space Science Institute, Boulder CO 80301 USA; 
email: {\tt travis@spsci.org}\\[\affilskip]
$^2$Stellar Astrophysics Centre, Aarhus University, DK-8000 Aarhus C, Denmark}

\pubyear{2013}
\volume{301}
\jname{Precision Asteroseismology}
\editors{W. Chaplin, J. Guzik, G. Handler \& A. Pigulski, eds.}
\begin{document}

\maketitle

\begin{abstract} 
Observations of magnetic activity cycles in other stars provide a broader 
context for our understanding of the 11-year sunspot cycle. The discovery 
of short activity cycles in a few stars, and the recognition of analogous 
variability in the Sun, suggest that there may be two distinct dynamos 
operating in different regions of the interior. Consequently, there is a 
natural link between studies of magnetic activity and asteroseismology, 
which can characterize some of the internal properties that are relevant 
to dynamos. I provide a brief historical overview of the connection 
between these two fields (including prescient work by Wojtek Dziembowski 
in 2007), and I highlight some exciting results that are beginning to 
emerge from the {\it Kepler} mission.


\end{abstract}

\firstsection
\section{Background}

In early 2005, I attended a seminar given by David Salabert in which he 
described his work documenting subtle shifts in the solar oscillation 
frequencies throughout the 11-year sunspot cycle \citep{sal04}. The 
measurements relied on data from the IRIS network, and they clearly showed 
that even the low-degree ($l\le3$) solar oscillation frequencies were 
shifted by a few tenths of a $\mu$Hz between magnetic minimum and maximum. 
If such frequency shifts were detectable in the Sun observed as a star, I 
realized that it might be possible to see similar behavior in other stars.

The high-degree oscillation frequency shifts in the Sun through the solar 
cycle were first characterized by \cite{lw90}. Using the thousands of 
oscillation modes then available from helioseismology they showed that the 
magnitude of the shift depended on both the geometry (spherical degree, 
$l$) and the frequency of the oscillation, with the largest shifts 
observed for higher degrees and at higher frequencies. The initial 
interpretation of these observations was given by \cite{gol91}, who 
matched the frequency dependence of the shifts by considering a direct 
magnetic perturbation to the near-surface propagation speed of the 
acoustic waves.

\cite{dg05} developed a similar formalism to explain modern space-based 
observations of the solar acoustic oscillations (p modes) as well as the 
surface gravity waves (f modes). They identified some secondary effects 
that were needed to explain the shifts observed in both sets of modes: a 
decrease in the radial component of the turbulent velocity, and the 
associated changes in temperature. Shortly after this work was published, 
I contacted Wojtek Dziembowski to ask whether I could use his code to 
calculate the expected shifts in low-degree p modes for other solar-type 
stars.

At the time there were very few stars with detections of solar-like 
oscillations, but \cite{fle06} would soon publish evidence of a marginally 
significant shift in the p mode frequencies of $\alpha$~Cen~A by comparing 
ground-based observations with earlier data from the WIRE satellite, and 
\cite{bed07} would see a similar (but statistically insignificant) shift 
when comparing two ground-based asteroseismology campaigns on $\beta$~Hyi. 
Wojtek happily sent me a copy of his code, and with his help I spent more 
than a year trying to figure out how to adapt it for other stars before he 
generously invited me to come to Warsaw for a week and work on it 
together. During that week we kick-started a project on $\beta$~Hyi, and 
we submitted the paper 6 months later \citep{met07}.

\section{Predictions}

As the activity level of the Sun rises from minimum to maximum, the p mode 
oscillations are gradually shifted to higher frequencies. The magnitude of 
the shift is proportional to the change in activity level, so the simplest 
prediction for other stars is to assume that the mean shift in the p mode 
frequencies scales with activity level. In other words, the largest shifts 
would be expected in the most active stars. This was the approach taken by 
\cite{cha07}, who were working contemporaneously.

Wojtek took a slightly different approach to predicting the frequency 
shifts in other stars. He parametrized the shifts as
$
\Delta\nu \propto A_0~[R/M]~Q_j(D_c),
$
where $A_0$ scales with the activity level as in \cite{cha07}, $R$ and $M$ 
are the radius and mass, and $Q_j$ is a function of $D_c$ which is the 
depth of the source of the perturbation below the photosphere. He 
demonstrated that this parametrization could remove all of the dependence 
on spherical degree and most of the frequency dependence in MDI data of 
the Sun with $D_c$ fixed at 0.3~Mm. To extend the relation to other stars, 
he assumed that $D_c$ would scale with the pressure scale height $H_p$ in 
the outer layers:
$
D_c \propto H_p \propto L^{1/4} R^{3/2} / M,
$
which can be expressed in terms of the luminosity $L$, radius and mass.

At the time we were doing this work, $\beta$~Hyi was the only star with a 
known activity cycle and multiple asteroseismic observing campaigns. The 
activity cycle had been observed in the Mg~{\sc ii} h and k lines by the 
IUE satellite, and several years of additional observations were available 
in the archive after the initial characterization by \cite{dra93}. Phil 
Judge did a complete reanalysis of the IUE data, and determined a cycle 
period of 12 years with a maximum at 1986.9. Marty Snow produced a 
comparable record of solar Mg~{\sc ii} h and k flux so we could scale the 
observed change in $\beta$~Hyi to predict a mean shift in the oscillation 
frequencies. Just by luck, $\beta$~Hyi was near magnetic maximum during 
the first detection of solar-like oscillations by \cite{bed01}, and close 
to magnetic minimum during the subsequent asteroseismic campaign by 
\cite{bed07}. The ground-based data were insufficient for a quantitative 
test, but Wojtek's relation qualitatively reproduced the observed shifts 
\citep{met07}.

\begin{figure}[t]
\begin{center}
\includegraphics[width=5.0in]{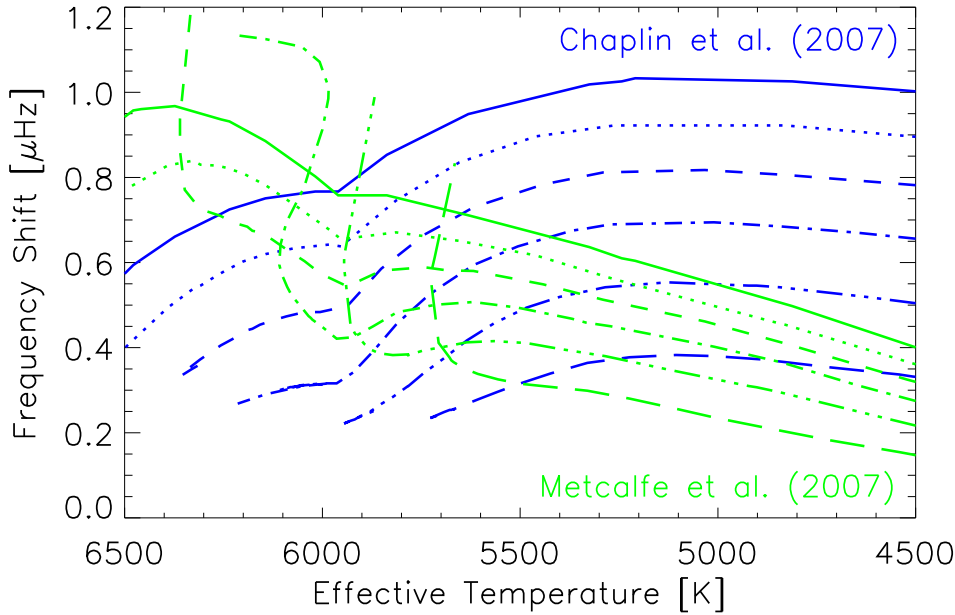} 
\caption{Predictions of the cycle-induced frequency shift as a function of 
effective temperature and age, with the scaling proposed by \cite{cha07} 
shown in blue and that proposed by \cite{met07} shown in green. Solid 
lines are Padova isochrones at an age of 1 Gyr, and the sets of dashed 
lines show progressively older isochrones at 1.6, 2.5, 4, 6.3, and 10 Gyr 
\citep[adapted from][]{kar09}.\label{fig1}}
\end{center}
\end{figure}

Two years later, Christoffer Karoff led a project to define an optimal 
sample of asteroseismic targets in the {\it Kepler} field that would also 
be monitored for stellar activity variations from ground-based Ca~{\sc ii} 
H and K measurements obtained throughout the mission \citep{kar09,kar13}. 
When trying to determine which stars would show the largest frequency 
shifts, he was confronted with two conflicting predictions (see 
Fig.~\ref{fig1}). The relation proposed by \cite{cha07} suggested that the 
largest shifts would be expected in the young active K stars. Wojtek's 
relation in \cite{met07} predicted that the hotter F stars would exhibit 
the largest shifts, which would grow even larger as the stars evolved. To 
be safe, the sample covered the full range of temperatures.

\section{Confirmation}

At the Beijing SONG workshop in March 2010, I gave a contributed talk on 
monitoring stellar magnetic activity cycles with SONG. During the coffee 
break Rafa Garc{\'{\i}}a told me about some asteroseismic observations of 
the F star HD\,49933, and he invited me to participate. The activity level 
of the star was not known, so I added it to a time-domain survey of 
Ca~{\sc ii} H and K emission that I was conducting for a sample of bright 
stars in the southern hemisphere \citep{met09}.

\cite{gar10} discovered anti-correlated changes in the frequencies and 
amplitudes of the oscillations in HD\,49933 during 150 days of continuous 
monitoring by CoRoT. The frequency shifts were positive, and passed 
through a minimum while the amplitudes increased and passed through a 
maximum---the same pattern of changes that occurs in the Sun as it passes 
through a magnetic minimum. Convection stochastically excites and 
intrinsically damps solar-like oscillations. Magnetic fields inhibit 
convection, suppressing the oscillation amplitudes while simultaneously 
shifting the frequencies.

\cite{sal11} pushed the analysis further, and examined the frequency 
dependence of the shifts. Just as in the Sun, the shifts grew steadily 
larger toward higher frequencies. It seemed clear that CoRoT had made the 
first asteroseismic detection of a stellar magnetic cycle, but there was 
one striking difference between HD\,49933 and the Sun. The frequency 
shifts observed in this $\sim$2~Gyr old F star at 6600~K were 4-5 times 
larger than solar, providing the first confirmation of Wojtek's relation.

\section{Future Prospects}

The archive of {\it Kepler} data represents an unprecedented opportunity 
to study the short-period magnetic cycles that have been observed in some 
rapidly rotating F stars. The high precision time-series photometry 
collected every 30 minutes over the past four years can be used to measure 
rotation periods from spot modulation and to monitor the longer-term 
brightness changes associated with the stellar cycle. Furthermore, for 
targets that have been observed in short-cadence (1-minute sampling), 
asteroseismology allows a characterization of the star including key 
dynamo ingredients such as the depth of the surface convection zone 
\citep{maz12} and radial differential rotation \citep{deh12}. The 
asteroseismic data can also be used to monitor the solar-like oscillations 
over time, allowing a search for the same pattern of changes that have 
been seen for the Sun and HD\,49933 in response to their magnetic cycles. 
Rapidly rotating F stars are the ideal targets because they show the 
shortest cycle periods \citep{met10}, and the frequency shifts are 
significantly larger than in the Sun (see Fig.~\ref{fig1}). Young, rapidly 
rotating K stars can also show relatively short cycles \citep{met13}, but 
the asteroseismic signatures are expected to be smaller.

\cite{mat13} recently examined the archive of {\it Kepler} observations 
for a sample of 22 rapidly rotating F stars. Wavelet analysis of the long 
light curves revealed clear signatures of latitudinal differential 
rotation and evidence for short magnetic cycles in a few stars. The best 
target in the sample has three years of continuous asteroseismic data 
spanning what appears to be a complete magnetic cycle, so additional tests 
of Wojtek's relation should soon be possible.

\begin{acknowledgments}

This long story has gradually unfolded with the collaboration and support 
of Tim Bedding, Phil Judge, Christoffer Karoff, Bill Chaplin, Rafa 
Garc{\'{\i}}a, Savita Mathur, and David Salabert. I would particularly 
like to thank Wojtek Dziembowski for graciously inviting me to work with 
him at the Copernicus Astronomical Center for a week in September 2006. 
This work was partially supported by NASA grant NNX13AC44G.

\end{acknowledgments}


\end{document}